# Availability-Aware and Efficiency-Driven AI Service Chain Provisioning in Multi-Domain Edge Intelligence Cloud

Hanzhi Chang, Jing Bai, Xin Tang, Xiaomei Liu, Yiming Chen

*Abstract*—**In a multi-domain edge intelligence cloud (MDEIC) managed by multiple network operators, AI services are delivered by chains of virtual network functions (VNFs) executed in sequence, called AI service chains (AISCs). Therefore, achieving an efficient and economical AISC provisioning approach is essential. However, the interaction between the environmental characteristics (heterogeneity, resource constraints and limited information visibility) of MDEIC and the time-dependence of AISCs, introduces various challenges to AISC provisioning in MDEIC. In this paper, we first formulate the AISC provisioning problem as a partially observable stochastic game (POSG). Then, we propose a graph-and-time-based multi-agent AISC provisioning (GT-MAAISCP) approach to achieve the collaborative optimization of AISC provisioning cost, delay and availability. Specifically, each agent uses the graph-time dueling network (GTDN) architecture to extract network topology information and temporal relationships. Finally, the experimental results demonstrate that the proposed approach outperforms benchmark approaches in MDEIC and also illustrate its performance under varying network topologies and different numbers of local EICs (LEICs).**

## I. Introduction

EDGE intelligence cloud (EIC) integrates mobile edge computing (MEC) and artificial intelligence (AI), which is expected to promote the vision of ubiquitous intelligence in 6G networks [1]-[3]. In EIC, the service nodes can run the inference and model training processes of energy- and compute-intensive, latency- and privacy-sensitive AI services, bringing AI closer to data and end devices [4]. Many service providers, including Google [5], Amazon [6] and Microsoft [7], have already launched EIC pilot programs. In real EIC, AI services such as autonomous driving and smart cities are provided by chains of virtual network functions (VNFs) executed in sequence, which are called AI service chains (AISCs) [8]. AISCs can provide users with efficient and flexible customized AI services.

EIC involves multiple types of resources, which are owned and managed by different network operators [9]. An AISC can be provisioned in multi-domain EIC (MDEIC) managed by multiple network operators. Fig. 1 shows an illustration of AISC provisioning in MDEIC. The AISC provisioning approach determines which EIC service nodes the VNFs run on and how the VNFs are interconnected on EIC service nodes. Therefore, it is important to carefully design the AISC provisioning approach to achieve efficient and economical selection of VNF locations and low-latency interconnection between VNFs. However, AISC provisioning faces the unresolved challenges found in traditional service function chain (SFC) provisioning, including heterogeneity, resource constraints, limited information visibility, and time-dependence. These challenges become more critical and complex due to the characteristics of AISCs and MDEIC. Unlike traditional SFC deployment [10], AISC provisioning focuses on AI-oriented challenges, such as cross-domain heterogeneity in management, satisfying specific resource constraints for AI, operating under limited information visibility, and handling complex time-dependent execution of sequential and parallel VNFs. In addition, AISCs tightly couple data locality, learning performance and service latency, which means that these challenges cannot be treated independently and must be jointly addressed in the provisioning process.

- Heterogeneity: To improve model accuracy, model training often requires access to data from multiple domains, which means AISC provisioning must be performed across multiple domains where the relevant data reside. This multi-domain deployment introduces a significant challenge of network heterogeneity, since the network topologies, node locations, and resource capacities vary across different domains. Ignoring such heterogeneity in MDEIC environments would reduce the adaptability and performance of AISC provisioning approaches.
- Resource constraints: AISCs must ensure both computational availability and model integrity. Continuous model updates introduce additional potential security risks, such as model poisoning or biased updates, which further increase the requirement for availability. EIC service nodes with limited resources are more susceptible to attacks [11][12]. Failures caused by these attacks can directly compromise AISC availability. Once availability is degraded, the accuracy of AI services also declines. Consequently, AISC provisioning faces more stringent resource constraints, requiring intelligent mechanisms to maintain reliability and security under limited resources.

Corresponding author: Jing Bai

Hanzhi Chang, Jing Bai, Xin Tang, Xiaomei Liu and Yiming Chen are with the School of Cyber Science and Engineering, University of International Relations, Beijing, China. E-mail: {hzchang, baijing, xtang, liuxiaomei, 20237103}@uir.edu.cn.

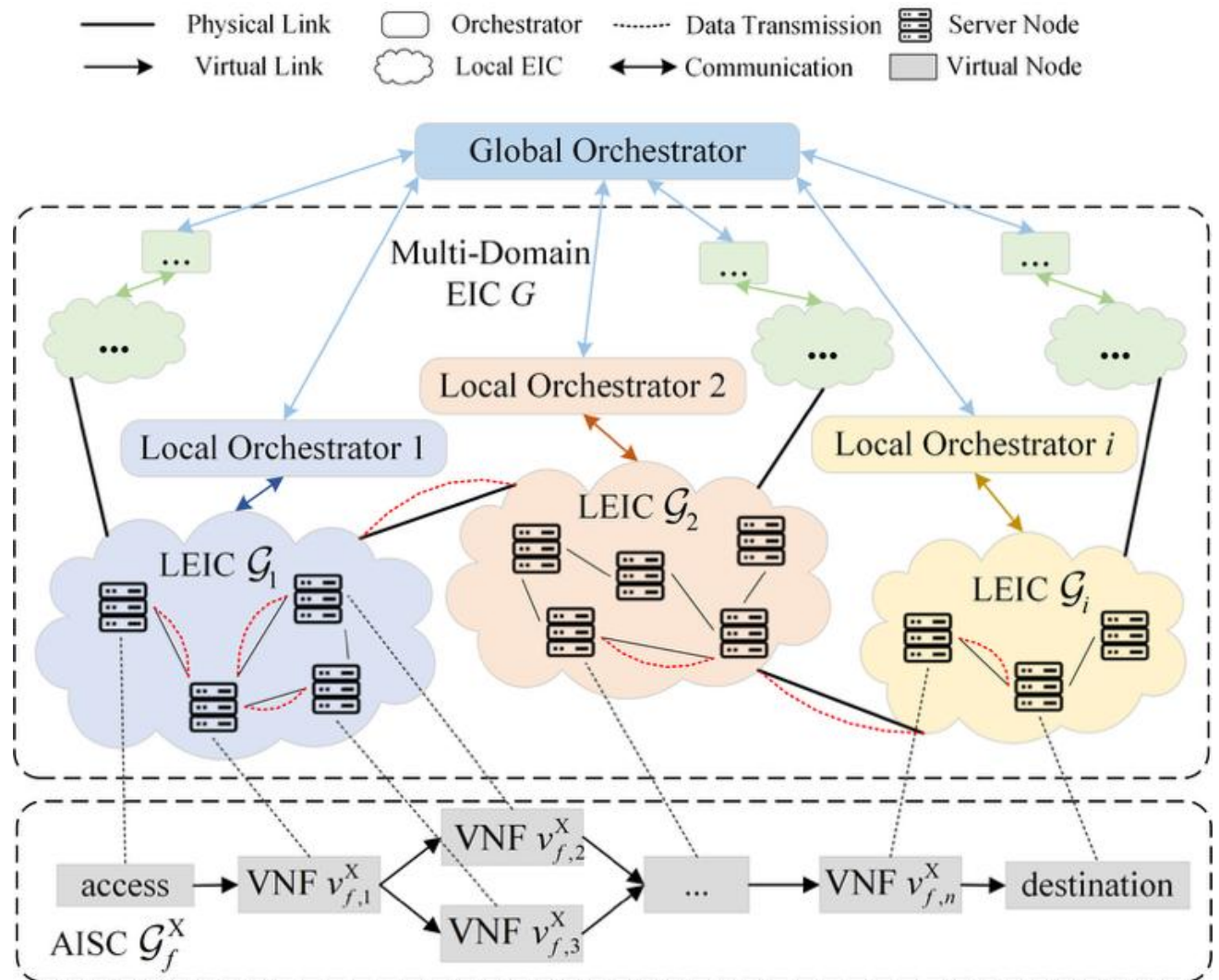

**Fig. 1.** The illustration of AISC provisioning in MDEIC

- Limited information visibility: In an MDEIC, different domains are managed by independent network operators. Training data collected within one domain cannot be freely shared with others, and private information such as physical topology and resource status is typically withheld across domains. This lack of transparency results in limited information visibility [9], making it more difficult to achieve efficient coordination and global optimization in AISC provisioning.
- Time-dependence: In AISCs, multiple VNFs exhibit may need to be executed either in sequence or in parallel [13]. Parallel VNFs often perform tasks such as feature extraction or data cleaning simultaneously before feeding their outputs to downstream inference VNFs, which introduces complex data flow synchronization requirements. If these parallel VNFs are deployed far apart or connected through low-bandwidth links, the latency advantages of parallel execution may be negated. Furthermore, multiple AISCs may arrive sequentially, making the consideration of such temporal relationships even more critical. Therefore, AISC provisioning approach must efficiently capture and handle the time dependencies among VNFs to ensure optimal performance.

Researchers have achieved fruitful results in terms of traditional service chain provisioning approaches, but *i*) existing heuristic approaches [14]-[20] lacked trial-and-error learning, which results in only local optimal solutions, and they were designed for specific problems, making it difficult to dynamically deploy VNFs and adapt to new physical network topologies; *ii*) existing single-agent deep reinforcement learning (DRL)-based approaches [21]-[25] had low learning efficiency, and they cannot process information from different domains simultaneously, making them inapplicable for service chain deployment in MDEIC; *iii*) existing multi-agent DRL (MADRL)-based approaches [26]-[32] fail to synergistically consider the impact of the environmental characteristics of MDEIC and the intrinsic properties of AISC on the AISC provisioning approach. As a result, existing studies are insufficient to support AISC provisioning in realistic MDEIC environments, especially when parallel VNFs, cross-domain coordination, and partial observability must be jointly considered.

To address the challenges mentioned above, we propose a graph-and-time-based multi-agent AISC provisioning (GT-MAAISCP) approach to achieve collaborative optimization of AISC provisioning cost, delay and availability. To the best of our knowledge, we are the first to address AISC provisioning problem in MDEIC, taking into account both the MDEIC environmental characteristics and the time-dependence of AISC. The contributions of this paper are summarized as follows:

- We formulate the AISC provisioning problem as a partially observable stochastic game (POSG) that addresses resource constraints and limited information visibility in MDEICs. In this formulation, multiple local agents (LAs) independently manage their respective local EICs (LEICs) under privacy and observability constraints, while a global agent (GA) coordinates cross-domain provisioning decision. Under four types of constraints (availability, delay, resource consumption and mapping constraints), each agent cooperates to complete AISC provisioning in MDEIC based on partial information.
- We design a Graph-Time Dueling Network (GTDN) architecture to address heterogeneity and time-dependence challenges. GTDN architecture enables agents to jointly capture graph and temporal dependencies in the provisioning process. The graph feature extractor effectively models heterogeneous network topologies and multi-dimensional link attributes, thereby addressing cross-domain heterogeneity. The temporal feature extractor captures interdependencies among VNFs and AISCs, allowing the system to handle both serial and parallel service chains. These components collectively enable adaptive decision-making in dynamic and time-dependent environments.
- We conduct extensive experiments under four representative network topologies—Watts-Strogatz Small-World, Barabási-Albert Scale-Free, Erdős-Rényi Random Graph, and Random Regular Graph. The results demonstrate that our approach achieves collaborative optimization of cost, delay, and availability under varying numbers of LEICs, significantly outperforming benchmarks on six performance metrics. These experiments confirm that the proposed approach maintains scalability, and multi-objective optimization capability in heterogeneous and dynamic MDEIC environments.

The rest of this paper is organized as follows. Section II introduces the related work. In Section III, we propose the system model and problem description. In Section IV, the graph-and-time-based multi-agent AISC provisioning approach is presented in detail. Section V shows the experimental results of verifying the performance of the proposed approach. Section VI concludes the paper and discusses future research directions.

## II. RELATED WORKS

In this section, we review the studies on service chain provisioning problem. There are two main solutions: heuristic approaches and DRL-based approaches. The latter can be further divided into single-agent DRL-based approaches and MADRL-based approaches. We detail these studies as follows.

### *A. Heuristic Approaches*

In the past few years, many studies tried to use heuristic approaches for optimizing service chain provisioning. Representative works have employed techniques such as column generation, mixed-integer linear programming, approximation algorithms, and metaheuristic optimization to improve deployment efficiency and resource utilization in NFV-enabled networks [14]–[18]. A detailed comparison of these approaches is summarized in Table I. In Table I, 'Multiple Domains' indicates whether a multi-domain EIC environment is considered. 'Parallel VNF' indicates whether the parallel VNF structures are included in service chains. 'Multiple Agents' indicates whether the MADRL algorithm is used for service chain provisioning. 'Topology Feature' indicates whether to use network topologies as a feature input to the MADRL algorithm. 'Temporal Feature' indicates whether the arrival order of service chains and the arrangement order of VNFs are utilized in the provisioning process.

Overall, heuristic approaches are generally designed for specific problems and have poor generalization ability, making it difficult to dynamically deploy VNF and adapt to new physical network topologies. Compared with heuristic-based approaches, when sufficient computational resources are available (e.g., CPU, memory, and GPU resources), the proposed approach achieves better generalization and adaptability to heterogeneous and dynamic environments. However, the RL-based nature of the proposed approach may introduce a certain degree of stochasticity during the training process, which can lead to variability in intermediate decision outcomes. To address this issue, our framework incorporates multiple constraints such as availability, delay, and resource consumption constraints. These constraints restrict the feasible action space and guide the agents to make stable and reliable provisioning decisions, thereby ensuring that the overall provisioning performance remains stable and suitable for real-world deployment scenarios.

### *B. DRL-based Approaches*

More and more studies use DRL algorithms for service chain provisioning. Representative works employed DRL techniques such as DQN, actor–critic architectures and graph-based models to optimize VNF placement and resource allocation under dynamic network conditions [21]–[25]. These approaches aimed to improve metrics such as acceptance ratio, long-term revenue and resource utilization. A detailed comparison of these methods is summarized in Table I.

However, the single-agent DRL-based approaches face significant limitations. Its learning efficiency and convergence speed is slow in large-scale EIC [33]–[35]. Additionally, it does not support cross-domain provisioning of service chains because it cannot process information from different domains simultaneously.

To address the aforementioned shortcomings, some studies have turned to the MADRL-based approaches to achieve better performance and generality when solving the service chain provisioning problem in large-scale or multi-domain environments [26]–[32]. In particular, Toumi *et al*. [27] proposed a MADRL framework based on DQN to address the challenge of deploying VNF in multi-domain environments. However, their model did not consider parallel VNF deployment, nor how agents utilize topological and temporal features to assist placement. The proposed graph-and-time-based multi-agent AISC provisioning (GT-MAAISCP) approach addresses these limitations. Wang *et al*. [32] proposed a GCN-based MADRL algorithm to solve the multi-objective optimization problem of dynamic VNF deployment in Internet of Things (IoT) scenarios. However, their method neglects edge features (e.g., bandwidth) in the network graph, and each agent is assigned to a single optimization objective, failing to solve the multi-domain collaboration. In contrast, our approach employs a more advanced graph neural network that can utilize multi-dimensional edge features in the topological network, resulting in superior representation capability and better performance.

Although studies on MADRL approaches for the service chain provisioning problem have made progress, there are still some limitations: *i*) Many studies [21]-[26][28][30]-[32] failed to address service chain provisioning problems in MDEIC and ignored the trustworthiness of the internal network, which affects approach scalability; *ii*) Some studies [24][26]-[30] ignored the impact of network topology on service chain provisioning, preventing these approaches from leveraging node location and link connectivity to optimize service chain provisioning; *iii*) Some studies [21]-[32] only focused on serial VNFs, and did not address the deployment of parallel VNFs; *iv*) Some studies [21]-[32] did not jointly optimize availability and other objectives, but the impact of availability on service chain provisioning cannot be ignored.

### *C. AISC Provisioning*

There are fewer studies on the AISC provisioning approach. Qiu *et al.* [19] proposed an online security-aware and reliability-guaranteed provisioning scheme for AISCs in edge intelligence clouds. Their model integrates security constraints and reliability requirements into a stochastic optimization framework to improve service stability. However, this approach relies on heuristic algorithm, leading to its limited adaptability to dynamic network states and large-scale scenarios and lacks the learning capability to optimize decisions over time. Moreover, this approach does not consider AISC provisioning across multiple domains. In terms of optimization objectives, it focuses solely on minimizing cost and improving security levels without considering delay performance. In contrast, our paper supports multi-domain AISC provisioning and jointly optimizes cost, delay, and availability through multi-agent reinforcement learning. Li *et al*. [8] presented a slicing-based AISC provisioning approach that balances AI service performance and data management resource consumption. This study focuses on single-domain environments and static resource slicing strategies. However, it does not consider cross-domain coordination. In addition, it only considered AISCs composed of serial VNFs and ignored AISCs containing parallel VNFs, which are common in AI-driven workloads. In contrast, our paper supports both serial and parallel VNF deployment in MDEIC through a graph-time dueling network architecture.

## III. System Model And Problem Description

In this section, we first introduce the system model in terms of the network model and the computation model. Then, we formulate the problem constraints and model the AISC provisioning problem in MDEIC as a partially observable stochastic game. Section A of the supplementary material introduces the notations used in this paper, their value in practical environments and representative real-world application examples.

### *A. Network Model*

As shown in Fig. 1, the MDEIC consists of multiple LEICs, each of which contains dozens or hundreds SNs. The server nodes (SNs) in each LEIC are interconnected by physical links (PLs), and the LEICs are also connected through inter-domain PLs. Each SN and PL is associated with specific attributes, such as bandwidth and availability. The network topology can be any type. An AISC comprises multiple VNFs that communicate with each other to deliver the intended service functions. These VNFs are deployed on SNs, and data are transmitted through the physical links between the SNs.

**Multi-Domain Edge Intelligence Cloud (MDEIC).** We model the MDEIC as a connected undirected graph $G = \{\mathcal{V}, \mathcal{E}\}$, which can be divided into $|G|$ subgraphs. Each subgraph denotes a local EIC (LEIC) $\mathcal{G}_i = \{\mathcal{V}_i, \mathcal{E}_i\}$, where $\mathcal{V}_i = \{v_{i,j}\}$ represents the set of service nodes (SNs) and $\mathcal{E}_i = \{e_{i,k}\}$ represents the set of physical links (PLs) between SNs ($1 \le i \le |G|, \mathcal{V}_i \in \mathcal{V}, \mathcal{E}_i \in \mathcal{E}$). A PL between two LEICs is represented by $e_{0,\mathcal{K}}$ ($e_{0,\mathcal{K}} \in \bar{\mathcal{E}} = \mathcal{E} \backslash \sum_{i=1}^{|G|} \mathcal{E}_i$, $1 \le \mathcal{K} \le \left|\mathcal{E} \backslash \sum_{i=1}^{|G|} \mathcal{E}_i\right|$, where \ is set difference). We set $|\mathcal{V}_i|$ SNs and $|\mathcal{E}_i|$ PLs in an LEIC $\mathcal{G}_i$, where the SN $v_{i,j} \in \mathcal{V}_i$ has resource capacity $C(v_{i,j})$ (i.e., computing resources, memory resources and storage resources), and the PLs $e_{i,k}$ and $e_{0,\mathcal{K}}$ have transmission bandwidth capacity, denoted as $C(e_{i,k})$ and $C(e_{0,\mathcal{K}})$, respectively ($1 \le j \le |\mathcal{V}_i|, 1 \le k \le |\mathcal{E}_i|$). Then, let $P(v_{i,j})$, $P(e_{i,k})$ and $P(e_{0,\mathcal{K}})$ be the unit resource costs for $v_{i,j}$, $e_{i,k}$ and $e_{0,\mathcal{K}}$, respectively. Furthermore, $a_{i,j}$ denotes the availability of $v_{i,j}$. $d_{i,k}$ and $d_{0,\mathcal{K}}$ denote the delay of $e_{i,k}$ and $e_{0,\mathcal{K}}$, respectively. Note that $i$, $j$ and $k$ are the LEIC index, SN index and PL index, respectively.

TABLE I
Comparison of the Existing Studies Discussed in Section II

| Reference | Physical Environment | | DRL Algorithm | | | Approach Performance Metrics | Optimization Objective |
|---|---|---|---|---|---|---|---|
| | Multiple Domains | Parallel VNF | Multiple Agents | Topology Feature | Temporal Feature | | |
| [12] | × | × | × | × | × | Bandwidth, Cost | Minimize the total costs |
| [13] | × | × | × | × | × | Delay, Bandwidth, Cost | Minimize the total costs, utilization, service interruption |
| [14] | × | × | × | × | × | Delay, Bandwidth, Cost, Reliability | Minimize the total costs, ensuring reliability and delay constraint |
| [15] | × | × | × | × | × | Delay, Bandwidth, Cost | Minimize the total costs and delay |
| [16] | × | × | × | × | × | Delay, Bandwidth, Cost | Optimize total costs and the acceptance ratio |
| [17] | × | × | × | × | × | Security level, Cost | Optimize security, the total costs, request throughput |
| [18], [20] | × | × | × | √ | × | Delay, Bandwidth, Cost | Minimize the total costs |
| [19] | × | × | × | √ | √ | Delay, Bandwidth, Cost | Optimize total costs and the acceptance ratio |
| [21] | × | × | × | × | × | Delay, Bandwidth, Cost | Maximize the acceptance ratio |
| [22] | × | × | × | √ | × | Delay, Bandwidth, Cost | Minimize the total costs |
| [23] | × | × | √ | × | × | Delay, Bandwidth, Cost | Maximize system utility |
| [24] | √ | × | √ | × | × | Delay, Bandwidth, Cost | Minimize costs and delay |
| [25] | × | × | √ | × | √ | Delay, Bandwidth | Minimize the total delay |
| [26] | √ | × | √ | × | × | Delay, Bandwidth, Cost | Minimize the total costs, delay and energy consumption |
| [27] | × | × | √ | × | √ | Delay, Cost | Minimize the total costs |
| [28] | × | × | √ | √ | × | Delay, Bandwidth, Cost | Minimize the total costs and acceptance ratio |
| [29] | × | × | √ | √ | × | Delay, Bandwidth | Minimize the total delay and resource utilization |
| This Paper | √ | √ | √ | √ | √ | Bandwidth, Delay, Cost, Availability | Optimize the total costs, delay and availability |

**Artificial Intelligence Service Chain (AISC).** An AISC consists of a series of VNFs, which are orchestrated in a given order. We set $\mathcal{G}_f^{\mathrm{x}} = \{\mathcal{V}_f^{\mathrm{x}}, \mathcal{E}_f^{\mathrm{x}}\}$ to define the $f$-th AISC, where $\mathcal{V}_f^{\mathrm{x}}$ denotes the set of VNFs and $\mathcal{E}_f^{\mathrm{x}}$ denotes the set of virtual links (VLs). The VNF $v_{f,n}^{\mathrm{x}} \in \mathcal{V}_f^{\mathrm{x}}$ has the resource requirement $C^{\mathrm{x}}(v_{f,n}^{\mathrm{x}})$ and availability $a_{f,n}^{\mathrm{x}}$. Two successive VNFs, $v_{f,n}^{\mathrm{x}}$ and $v_{f,n'}^{\mathrm{x}}$ are connected by a VL ($1 \le n, n' \le |\mathcal{V}_f^{\mathrm{x}}|$ and $n \neq n'$). Similarly, let $C^{\mathrm{x}}(e_{f,m}^{\mathrm{x}})$ be the bandwidth requirement of a VL $e_{f,m}^{\mathrm{x}} \in \mathcal{E}_f^{\mathrm{x}}$ ($1 \le m \le |\mathcal{E}_f^{\mathrm{x}}|$). We use $v_{f,i,j}^{\mathrm{in}}$ and $v_{f,i',j'}^{\mathrm{out}}$ to denote the access SN and destination SN of an AISC, respectively, which are specified by the AISC itself rather than determined by the provisioning approach. The purpose of setting AISC access and destination SNs is to allow users requiring cross-domain services to decide where to upload data and where to receive results. The provisioning approach is responsible for deploying AISC within the MDEIC under the given conditions of the specified access and destination SNs. In addition, $d_{\mathcal{G}_f^{\mathrm{x}}}$ denotes the maximum allowable link delay of an AISC and $a_{\mathcal{G}_f^{\mathrm{x}}}$ denotes the minimum allowable availability of an AISC. Note that $f$, $m$, and $n$ are the AISC index, VL index and VNF index, respectively.

**Remark**. By modeling an AISC as a directed graph, we can capture the behaviors of both sequential VNFs and parallel VNFs, which improves the generality of the approach.

*B. Computation Model*

To formulate delay, availability, cost and resource consumption, we use $\lambda$ as a binary variable. $\lambda_{v_{f,n}^{\mathrm{x}}, v_{i,j}}$ is a boolean value where if $\lambda_{v_{f,n}^{\mathrm{x}}, v_{i,j}} = 1$ then VNF $v_{f,n}^{\mathrm{X}}$ is deployed in server node $v_{i,j}$. $\lambda_{e_{f,m}^{\mathrm{x}}, e_{i,k}}$ is another boolean variable where if $\lambda_{e_{f,m}^{\mathrm{x}}, e_{i,k}} = 1$ then virtual link $e_{f,m}^{\mathrm{X}}$ is deployed in physical link $e_{i,k}$.

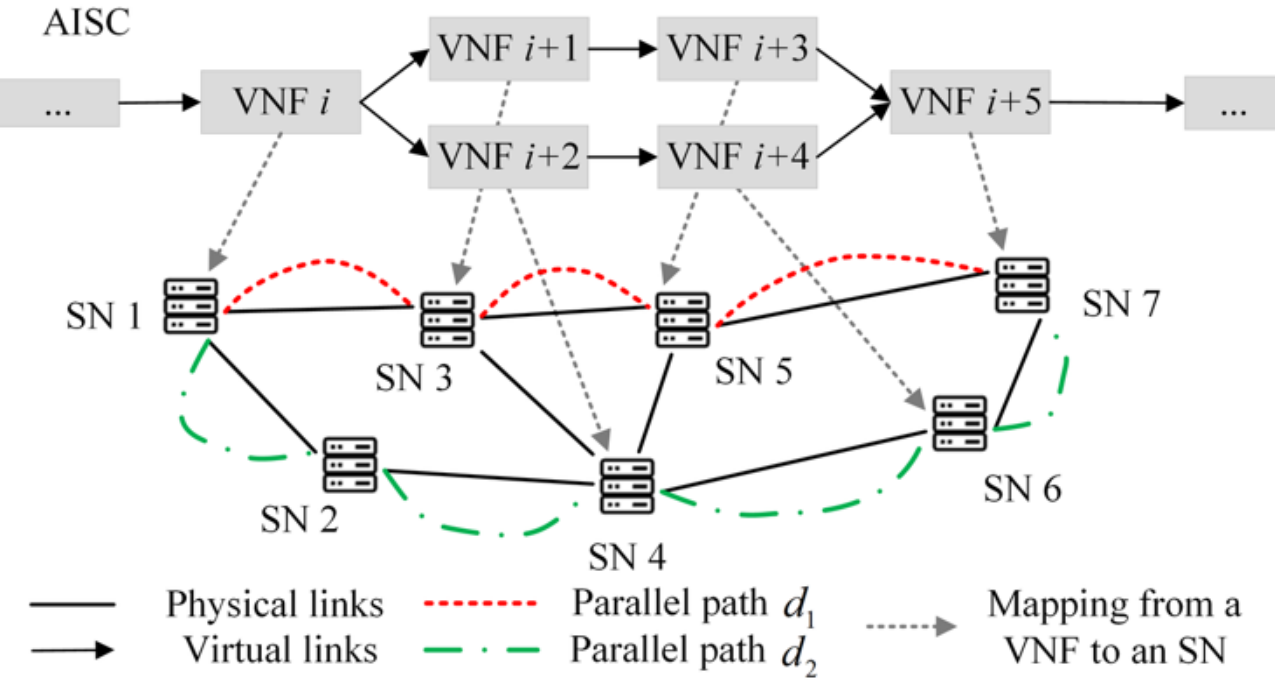

**Fig. 2.** The provisioning illustration of an AISC with two parallel paths

**Resource Consumption.** AISC provisioning in MDEIC consumes the resources of SNs and PLs. The resource consumption of each SN and PL can be calculated as follows,

$$\mathrm{Cons}_{v_{i,j}} = \sum_f \sum_{n=1}^{|\mathcal{V}_f^{\mathrm{x}}|} C^{\mathrm{x}}(v_{f,n}^{\mathrm{x}}) \cdot \lambda_{v_{f,n}^{\mathrm{x}}, v_{i,j}} \tag{1}$$

$$\mathrm{Cons}_{e_{i,k}} = \sum_f \sum_{m=0}^{|\mathcal{E}_f^{\mathrm{x}}|+1} C^{\mathrm{x}}(e_{f,m}^{\mathrm{x}}) \cdot \lambda_{e_{f,m}^{\mathrm{x}}, e_{i,k}} \tag{2}$$

$$\mathrm{Cons}_{e_{0,\mathcal{k}}} = \sum_f \sum_{m=0}^{|\mathcal{E}_f^{\mathrm{x}}|+1} C^{\mathrm{x}}(e_{f,m}^{\mathrm{x}}) \cdot \lambda_{e_{f,m}^{\mathrm{x}}, e_{0,\mathcal{K}}} \tag{3}$$

**Availability**. AISC availability refers to the probability that the AISC can operate without failure in the MDEIC. There are two scenarios for calculating availability: with backup and without backup. The former means that the AISC includes backup units for VNFs, while the latter does not. The AISC availability without backup can be calculated as follows,

$$\mathrm{Avail}_f = \prod_{i=1}^{|G|} \prod_{j=1}^{|\mathcal{V}_i|} \prod_{n=1}^{|\mathcal{V}_f^{\mathrm{x}}|} I\left(\lambda_{v_{f,n}^{\mathrm{x}}, v_{i,j}} \cdot a_{i,j} \cdot a_{f,n}^{\mathrm{x}}\right) \tag{4}$$

where $I(\cdot)$ is an indicator function that satisfies,

$$I(e) = \begin{cases} e, \text{if } e \neq 0 \\ 1, \text{if } e = 0 \end{cases} \tag{5}$$

The availabiilty of the AISC with backup is discussed in the Section.D of the supplementary materials.

*C. Problem Constraints*

In this section, we analyze the constraints when provisioning AISCs. The details are as follows.

**Delay Constraint**. AISC provisioning requires that transmission delay does not exceed the maximum allowable link delay to ensure the normal operation of the AISC. The delay constraint can be formulated as follows,

$$\forall \mathcal{G}_f^{\mathrm{x}}: \mathrm{Delay}_f \le d_{\mathcal{G}_f^{\mathrm{x}}} \tag{6}$$

**Resource Consumption Constraint.** Since the resources of SN are limited, the provisioning must ensure that it has sufficient capacity to run VNFs. The resource consumption constraint of SN can be formulated as follows,

$$\forall v_{i,j} \in \mathcal{V}_i: \mathrm{Cons}_{v_{i,j}} \le C(v_{i,j}) \tag{7}$$

Similarly, since PL bandwidth is limited, AISC provisioning must ensure that there is adequate available bandwidth. The resource consumption constraint for PL can be formulated as follows,

$$\forall e_{i,k} \in \mathcal{E}_i: \mathrm{Cons}_{e_{i,k}} \le C(e_{i,k}) \tag{8}$$

$$\forall e_{0,\mathcal{K}} \in |\bar{\mathcal{E}}|: \mathrm{Cons}_{e_{0,\mathcal{K}}} \le C(e_{0,\mathcal{K}}) \tag{9}$$

**Availability Constraints.** All components must be available, meaning that each SN hosting VNFs and each VNF in an AISC should be available. Therefore, the AISC availability constraint can be formulated as follows,

$$\forall \mathcal{G}_f^{\mathrm{x}}: \mathrm{Avail}_f \ge a_{\mathcal{G}_f^{\mathrm{x}}} \tag{10}$$

*D. Problem Formulation*

In this paper, we adopt a MADRL algorithm to deploy AISCs in MDEICs while ensuring privacy in each LEIC. The agent is the dispatcher who allocates resources to AISC. We classify agents into two types: local agent (LA) and global agent (GA). First, the GA decides to deploy VNFs to an LEIC. Then, the LA managing the LEIC deploys the VNF to an SN and PLs according to VLs. Considering privacy, the agent does not provide detailed information about its LEIC to other agents. Therefore, it is necessary to model the AISC provisioning process as a partially observable stochastic game (POSG) [37].

The POSG framework can capture both (1) the partially observable information of each agent's local environment and (2) the interactive and cooperative relationships among multiple agents in jointly optimizing global objectives such as cost, delay, and availability. By modeling AISC provisioning as a POSG, the proposed framework enables the agents to learn coordinated provisioning strategies that balance local optimization and global objectives in multi-domain environments. The following introduces how to formulate the AISC provisioning problem as a POSG, but the detailed design of POSG will be presented in Section IV-B and Section B in the supplementary materials.

In POSG, AISC provisioning starts at an initial state $s^0 \in S$. At each time $t$, the MDEIC is in state $s^t \in S$ with the previous joint provisioning action $a^{t-1} \in A$ (if $t = 0$, set $a^{t-1} = \emptyset$). The total number of agents is $|\mathcal{P}|$.The GA $p_g$ needs to take an action to choose an LEIC where a VNF is deployed. Next, the chosen LA $p_l$ needs to take an action to choose an SN for the final deployment of the VNF. Each agent $p$ has a policy $\pi_p(a_p^t|s_p^t)$ to take these actions. Given the provisioning action $a_p^t$, the environment transitions to the next state $s^{t+1} \in S$. Each agent participating in the provisioning receives a reward based on the reward function. These steps are repeated until completing a maximum number of $T$ time steps. As illustrated in Fig. 2, SN1–SN3 belong to LEIC 1, and SN4–SN6 belong to LEIC 2. When VNF $i$ is to be deployed, the system state is denoted as $s^{\mathrm{t}}$. The GA first decides to deploy VNF $i$ in LEIC 1. After receiving this instruction, the LA responsible for LEIC 1 determines to place VNF $i$ on SN1. As a result, the MDEIC state changes from $s^{\mathrm{t}}$ to $s^{\mathrm{t+1}}$ based on the actions of GA and LA. Then, the GA continues to select the placement of VNF $i+1$, and this process continues iteratively until the entire AISC is deployed.

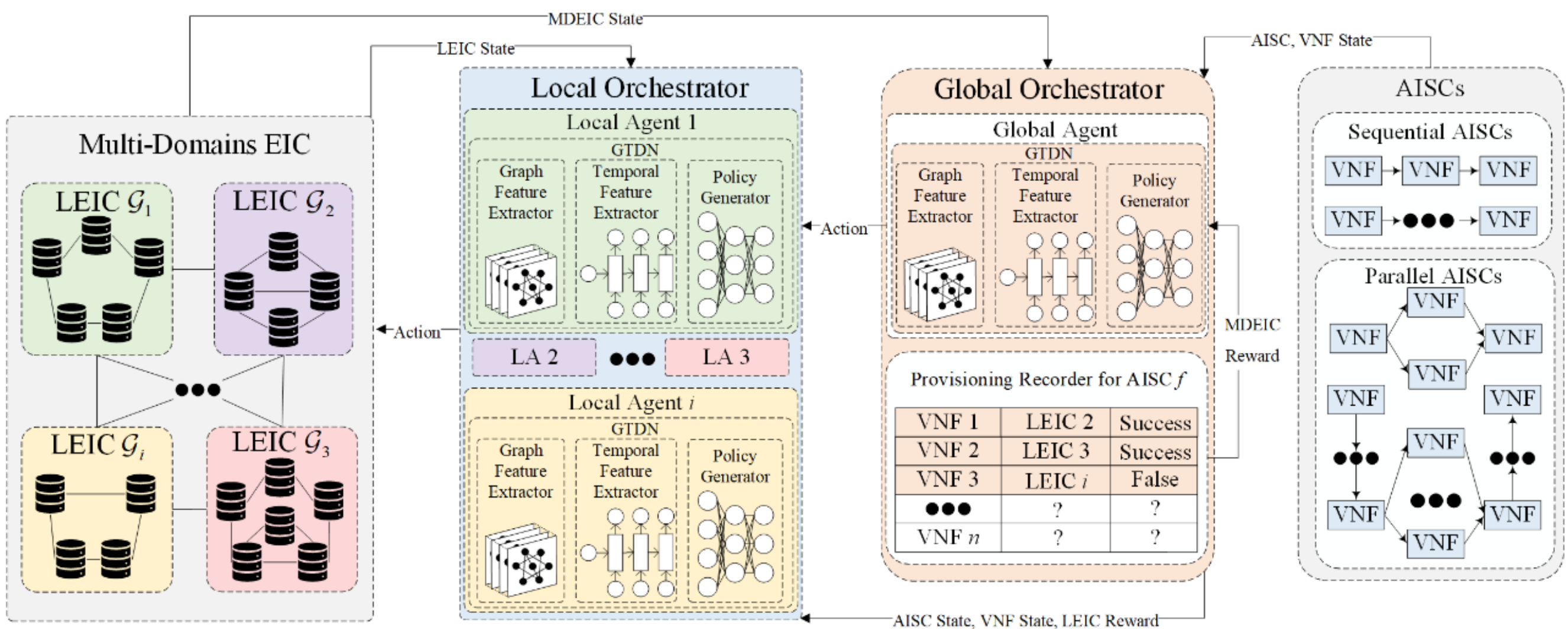

**Fig. 3.** The framework of GT-MAAISCP

## IV. Graph-and-Time-Based Multi-Agent AISC Provisioning Approach

In this section, we present an AISC provisioning approach using MADRL from three aspects: an explanation of the POSG model, a description of the graph-time dueling network (GTDN) architecture, and an elaboration on the learning process of this approach.

### *A. Overview*

We propose a graph-and-time-based multi-agent AISC provisioning approach (GT-MAAISCP), which can provision sequential AISCs and parallel AISCs in MDEIC. Fig. 3 illustrates the framework of GT-MAAISCP, where four components collaborate to perform AISC provisioning.

The MDEIC component is a large physical network composed of several LEICs. This component sends the states of the LEICs managed by the respective LAs to the local orchestrator and sends the states of the MDEIC to the global orchestrator.

The AISCs component is responsible for receiving various AISC requests, including sequential AISCs and parallel AISCs. This component sends the state of the VNF being deployed and the state of the AISC where the VNF is located to the global orchestrator.

The global orchestrator component includes the GA, which is responsible for assigning tasks to the local orchestrator, and the provisioning recorder, which tracks the actions of all agents. The GA takes an action based on the MDEIC state, VNF state and AISC state to determine on which LEIC the VNF should be deployed. It then sends the AISC state and VNF state to the LA managing the selected LEIC.

The local orchestrator component consists of many LAs. The LA takes an action based on the state of the LEIC it manages, the AISC state and VNF state received from the GA, to determine on which SN the VNF should be deployed. The provisioning recorder in the global orchestrator records the deployment results at each step. It calculates the rewards received by the GA and LA based on these results and sends them accordingly.

In the proposed approach, GA oversees multiple LAs in MDEIC, which is consistent with ETSI NFV architectural framework. That is, the GA corresponds to the NFVO in the ETSI NFV architectural framework, which serves as an orchestration platform responsible for coordinating VNFs, managing their life cycles, and handling multi-domain operations across heterogeneous infrastructures. Each LA corresponds to a VIM in the ETSI NFV architectural framework, which manages VNFs within its own LEIC [36][38].

In each agent, we propose a GTDN architecture that integrates the graph feature extractor (GFE), the temporal feature extractor (TFE) and the policy generator. Graph features refer to the structured representations that describe the relationships among nodes and edges in a graph-structured network, rather than treating them as isolated entities. In the AISC provisioning, the graph features extractor capture both the node attributes (e.g., resource capacity, availability, cost) and the edge attributes (e.g., bandwidth, delay) as well as their topological relationships. Temporal features refer to the sequential dependencies and temporal correlations among VNFs in an AISC. These features describe how the state evolves over time and how previous deployment decisions influence subsequent decisions. In the AISC provisioning, the temporal features extractor capture information such as the deployment order of VNFs, dynamic changes in performance (e.g., delay accumulation) and historical placement relationships. For example, when deploying VNF $i+2$, the delay constraint may depend not only on the immediately preceding VNF $i+1$ but also on the earlier VNF $i$, especially when parallel branches are involved. Residual gated graph convolutional network (RGGC) is used in the GFE to extract graph features from physical networks and AISCs. Gated recurrent units (GRU) are used in the TFE to extract temporal features of AISCs and VNFs in an AISC. We utilize a dueling network as a policy generator, which separates the policy generation network into a value network and an advantage network.

**Remark 1.** Since the MDEIC, LEICs and AISCs are represented by different graphs, RGGC can effectively capture node attributes and edge attributes. Moreover, AISCs arrive sequentially, and the VNFs in each AISC are deployed in order, making GRU suitable for extracting temporal features of AISCs and their VNFs. Its memory capability further enhances its ability to handle sequential input effectively. The dueling network enables the MADRL algorithm to reduce redundant information in the learning process, improve learning efficiency and enhance policy stability.

**Remark 2.** AISC provisioning is performed sequentially at the granularity of VNFs. To satisfy service delay constraints, adjacent VNFs should ideally be placed on nearby SNs. Therefore, during the sequential VNF deployment, the agent must remember the characteristics and placement of previously deployed VNFs to make optimal decisions for subsequent ones. Moreover, for an AISC containing parallel VNFs, the agent must capture longer-term dependencies between non-adjacent VNFs. For instance, as shown in Fig. 2, when the deployment order is VNF $i$, VNF $i+1$ and VNF $i+2$, the delay of VNF $i+2$ is primarily influenced by the placement of VNF $i$ rather than VNF $i+1$. Therefore, employing GRU as the temporal feature extractor is essential for learning these temporal dependencies and improving delay performance.

**Remark 3.** The 'divide and conquer' strategy of MADRL algorithm enables LEIC providers to protect network privacy and enhance provisioning capabilities in large-scale MDEIC environments. LAs can be trained independently, and LEIC providers are not required to share internal information. The GA receives only aggregated data from each LEIC, which means that providers only need to provide high-level summaries of their networks without disclosing detailed internal information. The GA is trained through the jointly learned policies submitted by all participating LAs, without relying on privileged access to any specific domain. Moreover, the network capacity can be expanded by increasing the number of LAs without affecting the existing LAs.

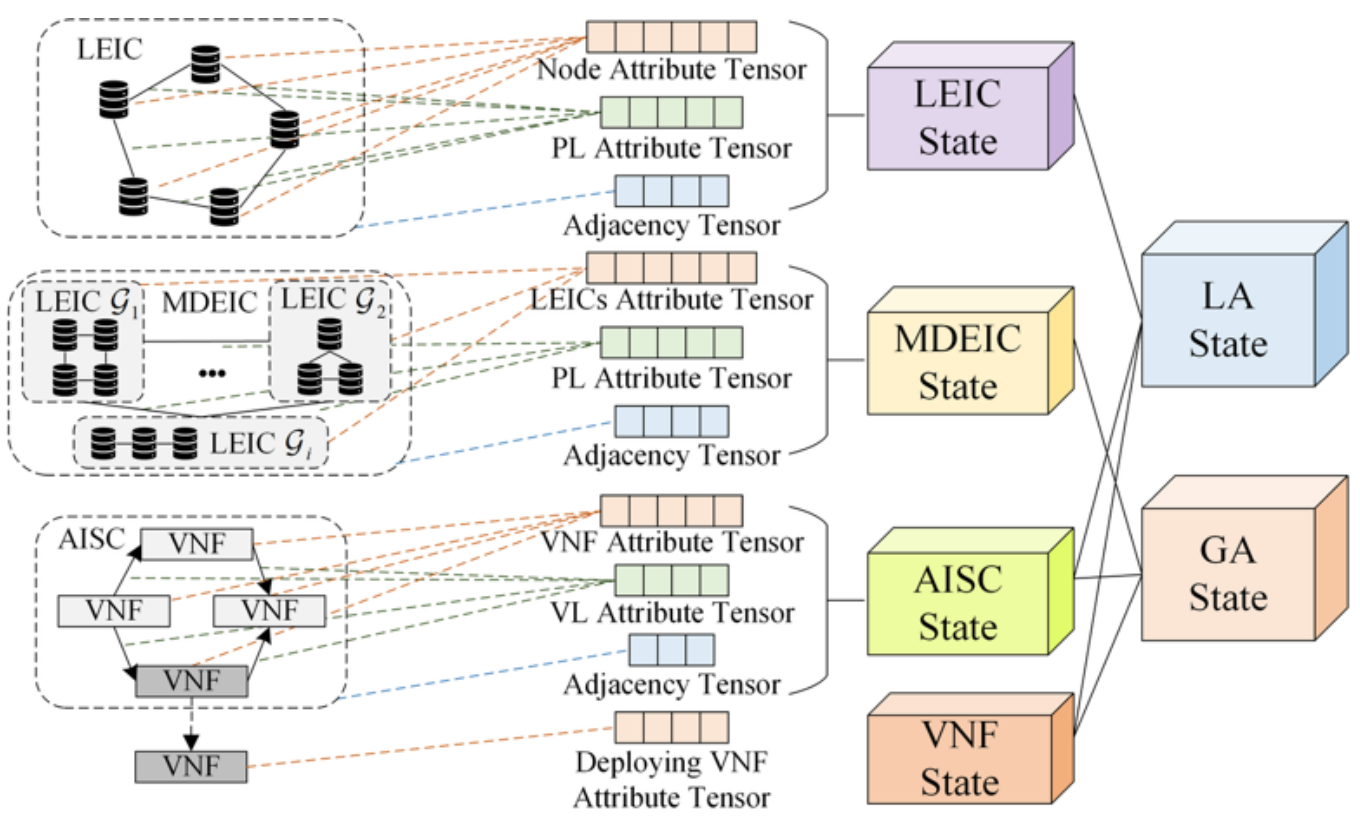

**Fig. 4.** The state of the agents

### *B. State, Action and Reward Representation*

**Action.** At each decision step, the GA selects one LEIC from the LEIC set to deploy the current VNF. Subsequently, the corresponding LA selects a specific SN within that LEIC to host the VNF. The actions of the GA and LA are shown as follows,

$$A_{\text{global}} \in [1, |G|] \tag{11}$$

$$A_{\text{local},i} \in [1, |\mathcal{V}_i|] \tag{12}$$

The size of the action space of GA $|A_{\text{global}}|$ corresponds to the number of LEICs, while the size of the action space of the $i$-th LA $|A_{\text{local},i}|$ is equal to the number of SNs within its managed LEIC.

### *C. Graph-Time Dueling Network Architecture*

The graph-time dueling network (GTDN) architecture is illustrated in Fig. 5 and consists of four components: *i*) Physical network state feature extraction; *ii*) AISC state feature extraction; *iii*) VNF state feature extraction; *iv*) Policy generator. Each component is explained in more detail as follows.

**Policy generator.** In our approach, we use dueling network architectures for DQN to generate the policy. For an agent following policy $\pi$, the values of action $a$ and state $s$ are defined as follows,

$$Q^{\pi}(s,a) = \mathbb{E}_{\pi}[R_t | s_t = s, a_t = a] \tag{13}$$

$$V^{\pi}(s) = \mathbb{E}_{a\sim\pi(s)}[Q^{\pi}(s,a)] \tag{14}$$

The value function $V^{\pi}(s)$ quantifies the expected value of being in a given state $s$, while $Q^{\pi}(s,a)$ assesses the expected value of taking a specific action in state $s$. To approximate the state-action value function $Q^{\pi}(s,a)$, we utilize a deep Q-network: $Q(s,a;\theta)$ with parameter $\theta$. The loss function of the network at the $i$-th iteration is defined as follows,

$$L_i(\theta_i) = \mathbb{E}_{s,a,r,s'\sim\mathcal{U}(\mathcal{D})}\left[\left(y_i^{DQN} - Q(s,a;\theta_i)\right)^2\right] \tag{15}$$

$$y_i^{DQN} = r + \gamma \max_{a'} Q(s',a';\theta^-) \tag{16}$$

where $\theta^-$ represents the parameters of the fixed target network. During training, the agent maintains a dataset $\mathcal{D}_t = \{E_1, \cdots, E_t\}$, where each experience $E_t = (s_t, a_t, r_t, s_{t+1})$ is collected from multiple episodes.

The dueling network architecture independently estimates the state-value function $V(s)$ and the advantage function $A(s,a)$ without additional supervision. The advantage function is defined as follows:

$$A^\pi(s,a) = Q^\pi(s,a) - V^\pi(s) \tag{17}$$

To aggregate these functions, we use the following module:

$$Q(s,a;\theta,\alpha,\beta) = V(s;\theta,\beta) + \left(A(s,a';\theta,\alpha) - \frac{1}{|\mathcal{A}|}\sum\nolimits_{a'} A(s,a';\theta,\alpha)\right) \tag{18}$$

where $\alpha$ and $\beta$ are the parameters of two separate outputs of neural networks. The formula also stabilizes the optimization by requiring the advantages at a rate consistent with the mean value. The output of the state-action value is a vector, where each element represents a specific SN managed by an LA or an LEIC managed by the GA.

*D. Multi-agent Reinforcement Learning Approach*

In our approach, we use independent learning to train agents. Each LA $p_l$ learns its own policy $\pi_{p_l}$ based solely on its own experiences, without considering other agents. The effects of other agents' actions are treated as part of the environment dynamics. In our state design, each LA's state is not affected by others, thus avoiding the primary drawback of independent learning, namely, non-stationarity caused by concurrent learning among multiple agents. Additionally, independent learning often produces results comparable to state-of-the-art MADRL algorithms, as shown in [39].

The the AISC provisioning method as shown in Alg. 2, respectively. If an agent selects an action that violates the defined constraints, a penalty reward is assigned. Otherwise, when the selected action satisfies the constraints, the agent continues the decision-making process until the entire AISC is successfully deployed. To calculate the delayed reward after AISC provisioning is complete, we temporarily store the experience tuple $(s, a, 0, s')$ from each VNF deployment in a buffer and use a reward value of 0 to each experience. Once the AISC provisioning is complete, we update this buffer with the actual reward values and integrate these experiences into the agent's dataset. If early VNFs are placed improperly and cause subsequent deployment failures, the corresponding experiences receive lower or negative rewards, thereby guiding the agent to refine its strategy in future episodes.

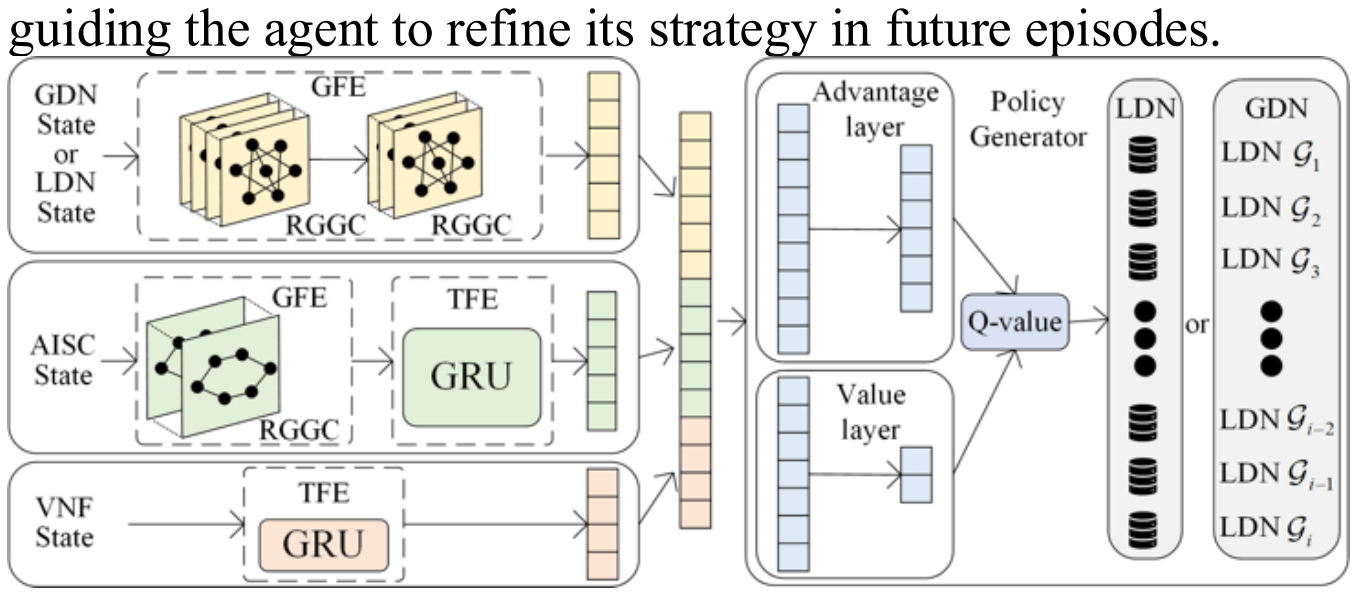

**Fig. 5.** The graph-time dueling network architecture

In addition, the GT-MAAISCP framework inherently addresses the problem of deploying neighboring VNFs on distant service nodes. As described in Alg. 1, when a new AISC arrives, the GA first decides which LA will handle the deployment of the first VNF. If the GA allocates two neighboring VNFs to distant LAs, it will cause excessive inter-node delay, eventually leading to deployment failure or significantly reduction in reward, as defined in Eq. (29) and Alg. 2. Through reinforcement learning, the GA gradually learns that this allocation is not optimal, so it adjusts its strategy to deploy neighboring VNFs closer together to minimize latency. The same principle applies to the LAs. That is, if the GA assigns multiple VNFs to the same LA consecutively, the LA can also learn from the reward mechanism to place neighboring VNFs on neighboring SNs to minimize intra-domain delay. In Alg. 2, we use Dijkstra algorithm to identify the shortest path between SNs decided by two successive actions.

If the VNF or VL placement satisfies the constraints, it is successfully allocated to an SN or a PL. If a VNF fails to deploy, all previously allocated physical resources of the corresponding AISC are released, and the agent receives a penalty. This penalty feedback encourages the agent to avoid infeasible deployment decisions and gradually learn strategies that improve the success rate of AISC provisioning.

## V. Performance Evaluation and Analysis

This section presents simulation results to verify the performance of the proposed GT-MAAISCP approach. The experiments are implemented by using Pytorch [40] and NetworkX, conducted on a computer with an AMD Ryzen 9 7945HX CPU, an RTX 4060 GPU and 64 GB RAM.

*A. Performance Evaluation*

**Ablation Experiment:** We adopt representative approaches from existing related studies to conduct the ablation experiments, rather than directly removing individual modules from our framework. In the first comparison experiment (labeled as Approach 1), we implement the approach proposed by Wang *et al*. [22], which replaces RGGCs in GFE with GCN, while preserving the TFE configuration. In the second comparison experiment (labeled as Approach 2), we implement the approach proposed by Xu *et al*. [25], which replaces both GFE and TFE with MLP layers. In the third comparison experiment (labeled as Approach 3), we adopt the method of Wang *et al*.[32], in which the TFE module is removed, and GCN is used as GFE. These three benchmark approaches are derived from existing published studies, allowing us to systematically evaluate the contribution of each architectural component (GFE and TFE) within our framework under consistent experimental conditions.

Fig. 6 (a) illustrates the accumulated reward obtained by different methods as the number of training episodes increases. After convergence, the proposed approach achieves the highest reward value (exceeding 400), while the convergence range of baseline methods was approximately 200 to 400. In addition, Fig. 6 (b) presents the variation of the loss values with the number of training updates. Compared with Approach 2 (using MLP), the proposed approach exhibits lower loss fluctuations and more stable convergence. Furthermore, compared with Approach 3 (without the TFE module) and Approach 1 (without RGGC), the proposed approach achieves lower loss values, demonstrating the effectiveness of integrating both graph-based and temporal feature extraction modules.

Fig. 7 illustrates the performance of different approaches in different metrics. All GFE-based approaches (Approach 1, 3 and GTDN) outperform the non-GFE approach (Approach 2) on most metrics. Although MLP can handle unstructured data effectively, it cannot capture node relationships in graph-structured data. However, in terms of AISC availability, Approach 1 is about 1% worse than Approach 2. Combining GCN with GRU does not fully exploit the advantages of both components, which results in inadequate temporal dependency modeling and inefficient information propagation. In contrast, our approach demonstrates the compatibility of combining RGGC with GRU, achieving approximately 1.6% higher AISC availability than the non-TFE approach (Approach 3).

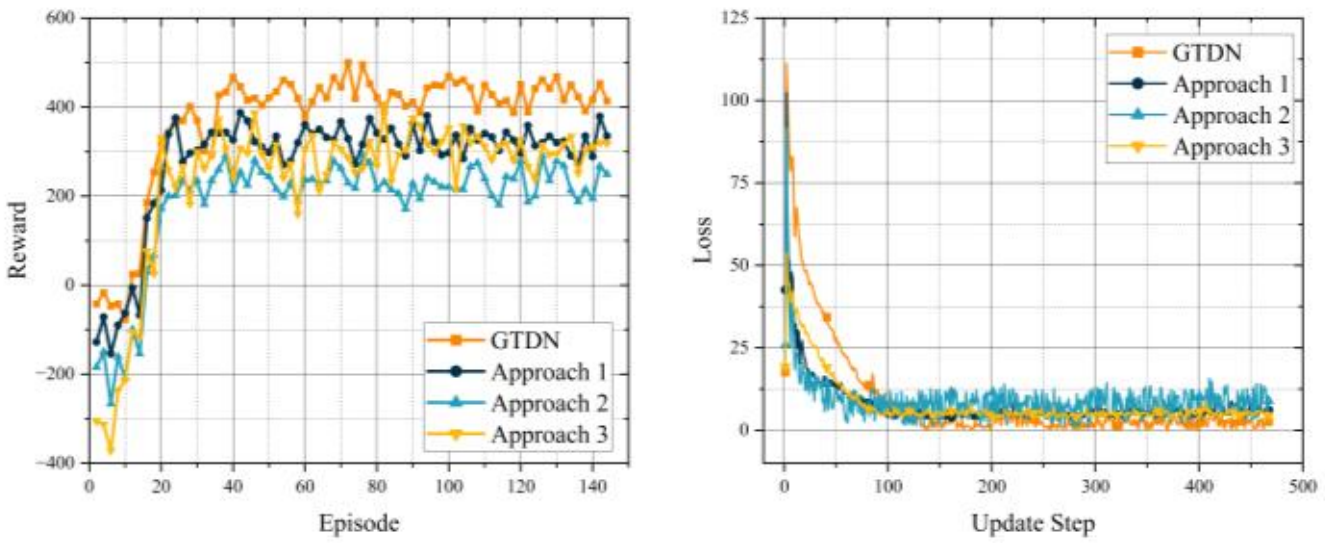

**Fig. 6.** The training performance of different approaches

We also find that the number of PLs traversed by AISC is positively correlated with the PL bandwidth costs. These results indicate that our agent can reduce bandwidth costs by optimizing the physical position of VNFs. We can conclude that incorporating link attributes into our approach enhances its applicability for the AISC provisioning problem compared to GCN without using link attributes. For example, in our approach, an AISC initially spans about 10-12 PLs, but it converges to 2-4 PLs after training, which means that transmission delay is reduced and some VNFs are deployed in the same SN to reduce bandwidth costs. Overall, the GTDN architecture demonstrates more effective performance than other comparison approaches.

Note that when conducting the ablation experiments, we select the DQN with a dueling network as the policy generator, merely as an example to validate the effectiveness of the proposed framework [41]. Since the RL component in our framework is implemented in a modular manner, which allows various advanced RL algorithms (e.g., DQN, A3C, PPO) to be seamlessly integrated into the same architecture, we can also select other algorithms for experiments. Regardless of which RL algorithm is adopted, the experimental results will lead to the same conclusion that the proposed framework effectively enhances AISC provisioning performance through the integration of graph-based and temporal feature extraction.

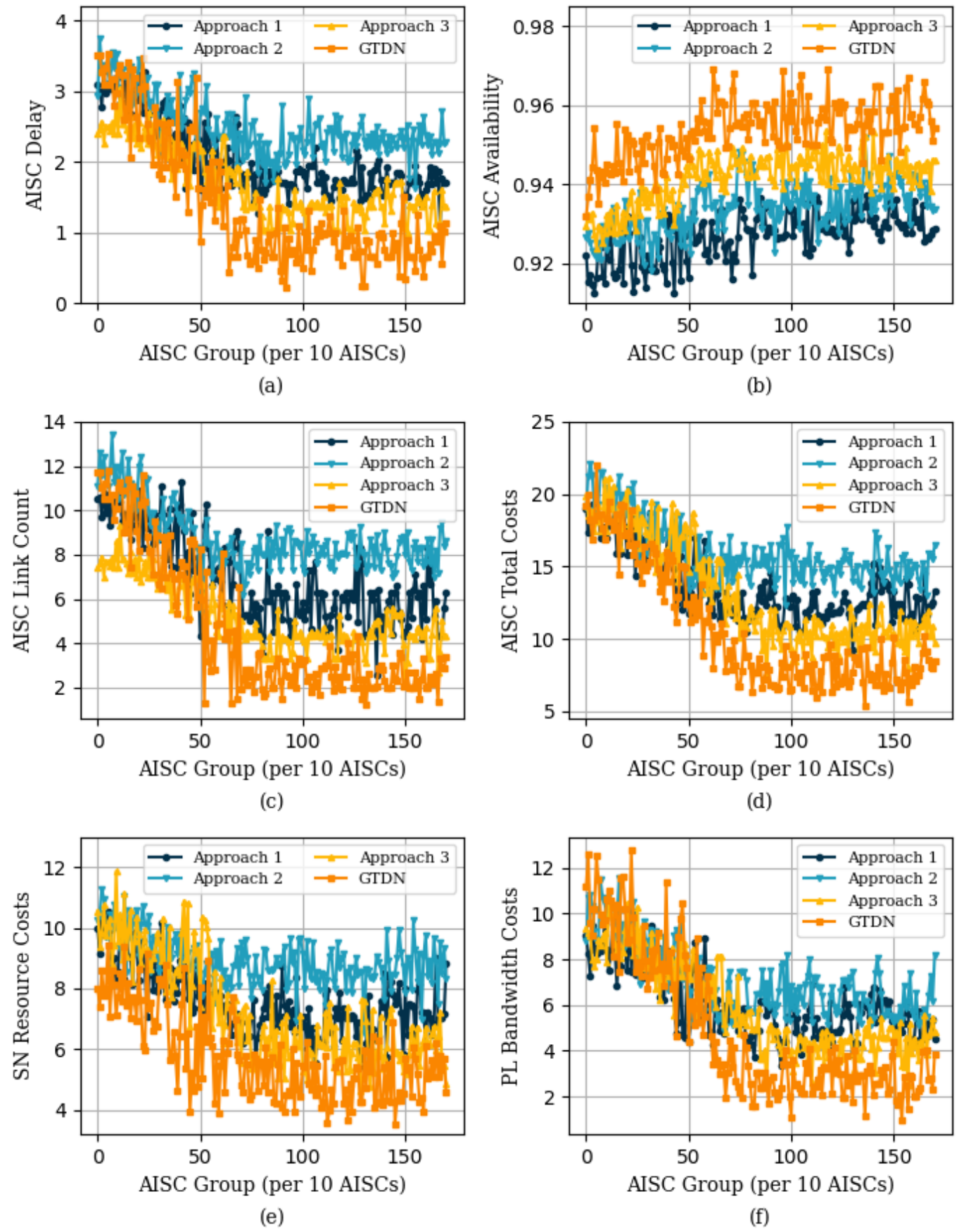

**Fig. 7.** The performance of different approaches in different metrics

## VII. Conclusion

In this paper, we investigate the AISC provisioning problem in MDEIC, considering both the MDEIC environmental characteristics and the time-dependence of AISC. First, we establish the network model and computation model of MDEIC, considering multiple factors such as bandwidth, delay, costs and availability. Then, based on POSG model, we transform the AISC provisioning problem into a multi-objective optimization problem under multiple constraints. To solve this problem, we propose the GT-MAAISCP approach to achieve collaborative optimization of AISC provisioning cost, delay and availability. The experimental results demonstrate the effectiveness of the proposed approach for provisioning AISCs in MDEIC.

Nevertheless, several challenges remain, including scalability in large-scale MDEIC environments, the design of reliable backup strategies, and the adaptability of the framework to highly dynamic network conditions. In future work, we will further investigate robust backup mechanisms to enhance adaptability under highly dynamic and non-stationary conditions, and validate it in real-world MDEIC environments.

## References

[1] Hemmati, P. Raoufi, and A. M. Rahmani, “Edge artificial intelligence for big data: a systematic review,” *Neural Comput & Applic*, vol. 36, no. 19, pp. 11461–11494, 2024.

[2] X. Wang, B. Wang, Y. Wu, Z. Ning, S. Guo, and F. R. Yu, “A Survey on Trustworthy Edge Intelligence: From Security and Reliability To Transparency and Sustainability,” *IEEE Communications Surveys & Tutorials*, vol.27, no. 3, pp. 1729–1757, 2025.

[3] H. Cao, M. Alrashoud, T. Mohamed, and L. Yang, “An Intelligent Softwarized Resource Management and Allocation Framework for Services With Personalized Intentions in 6G-Enabled IoT Networks,” *IEEE Internet of Things Journal*, vol. 13, no. 5, pp. 8261–8274, 2026.

[4] Z. Zhou, X. Chen, E. Li, L. Zeng, K. Luo, and J. Zhang, “Edge Intelligence: Paving the Last Mile of Artificial Intelligence With Edge Computing,” *Proceedings of the IEEE*, vol. 107, no. 8, pp. 1738–1762, 2019.

[5] Google, “Google AI Edge,” [Online]. Available: https://ai.google.dev/edge, 2025.

[6] E. Mercado, R. Shira, and C. Samynathan, “Unlocking the Power of Edge Intelligence with AWS,” [Online]. Available: https://aws.amazon.com/blogs/iot/unlocking-real-time-intelligence-at-the-edge-with-awss-connected-edge-intelligence/, 2024.

[7] V. Bahl, M. Chiang, U. Ramachandran, A. Samuel, L. Zhong, and K. Jameson, “Intelligent Edge,” [Online]. Available: https://www.microsoft.com/en-us/research/video/intelligent-edge/, 2018.

[8] M. Li, J. Gao, C. Zhou, X. S. Shen, and W. Zhuang, “Slicing-Based Artificial Intelligence Service Provisioning on the Network Edge: Balancing AI Service Performance and Resource Consumption of Data Management,” *IEEE Vehicular Technology Magazine*, vol. 16, no. 4, pp. 16–26, 2021.

[9] W. Liu, N. Hua, X. Zheng, and B. Zhou, “Intelligent inter-domain connection provisioning for multi-domain multi-vendor optical networks,” *Journal of Optical Communications and Networking*, vol. 7, no. 3, pp. 176–192, 2015.

[10] X.-Q. Xi and Y. Zhang, “A Construction Mechanism of Routing Service Chain Based on the Service Customization,” in *2015 International Conference on Computer Science and Applications* (CSA), 2015, pp. 318–322.

[11] Q. He *et al.*, “A Game-Theoretical Approach for Mitigating Edge DDoS Attack,” *IEEE Transactions on Dependable and Secure Computing*, vol. 19, no. 4, pp. 2333–2348, 2022.

[12] Q. Wang, S. Zou, Y. Sun, M. Liwang, X. Wang, and W. Ni, “Toward Intelligent and Adaptive Task Scheduling for 6G: An Intent-Driven Framework,” *IEEE Transactions on Cognitive Communications and Networking*, vol. 10, no. 5, pp. 1975–1988, 2024.

[13] X. Yu *et al.*, “Priority-Aware Deployment of Autoscaling Service Function Chains Based on Deep Reinforcement Learning,” *IEEE Transactions on Cognitive Communications and Networking*, vol. 10, no. 3, pp. 1050–1062, 2024.

[14] J. Liu, W. Lu, F. Zhou, P. Lu, and Z. Zhu, “On Dynamic Service Function Chain Deployment and Readjustment,” *IEEE Trans. Netw. Serv. Manage.*, vol. 14, no. 3, pp. 543–553, 2017.

[15] R. Behravesh, D. Harutyunyan, E. Coronado, and R. Riggio, “Time-Sensitive Mobile User Association and SFC Placement in MEC-Enabled 5G Networks,” *IEEE Trans. Netw. Serv. Manage.*, vol. 18, no. 3, pp. 3006–3020, 2021.

[16] V. R. Chintapalli, B. R. Killi, R. Partani, B. R. Tamma, and C. S. R. Murthy, “Energy- and Reliability-Aware Provisioning of Parallelized Service Function Chains With Delay Guarantees,” *IEEE Trans. on Green Commun. Netw.*, vol.8, no. 1, pp. 205–223, 2023.

[17] M. Asgarian, K. Jamshidi, and A. Bohlooli, “An Efficient Approximation Algorithm for Service Function Chaining Placement in Edge–Cloud Computing Industrial Internet of Things,” *IEEE Internet Things J.*, vol. 11, no. 7, pp. 12815–12822, 2024.

[18] S. R. Zahedi and S. Jamali, “A hybrid model for VNF deployment capable of responding to online requests at network edge,” *Computer Networks*, vol. 247, p. 110385, Jun. 2024, doi: 10.1016/j.comnet.2024.110385.

[19] Y. Qiu, J. Liang, V. C. M. Leung, and M. Chen, “Online Security-Aware and Reliability-Guaranteed AI Service Chains Provisioning in Edge Intelligence Cloud,” *IEEE Trans. on Mobile Comput.*, vol. 23, no. 5, pp. 5933–5948, 2024.

[20] X. Li, H. Zhou, L. Ma, J. Xin, and S. Huang, “Cost and Latency Customized SFC Deployment in Hybrid VNF and PNF Environment,” *IEEE Transactions on Network and Service Management*, vol. 21, no. 4, pp. 4312–4331, 2024.

[21] J. Pei, P. Hong, M. Pan, J. Liu, and J. Zhou, “Optimal VNF Placement via Deep Reinforcement Learning in SDN/NFV-Enabled Networks,” *IEEE J. Select. Areas Commun.*, vol. 38, no. 2, pp. 263–278, Feb. 2020.

[22] T. Wang *et al.*, “DRL-SFCP: Adaptive Service Function Chains Placement with Deep Reinforcement Learning,” in *ICC 2021 - IEEE International Conference on Communications*, 2021, pp. 1–6.

[23] H. Liu, S. Ding, S. Wang, G. Zhao, and C. Wang, “Multi-objective Optimization Service Function Chain Placement Algorithm Based on Reinforcement Learning,” *J Netw Syst Manage*, vol. 30, no. 4, p. 58, 2022.

[24] L. Yang, J. Jia, H. Lin, and J. Cao, “Reliable Dynamic Service Chain Scheduling in 5G Networks,” *IEEE Trans. on Mobile Comput.*, vol. 22, no. 8, pp. 4898–4911, 2023.

[25] J. Xu, X. Cao, Q. Duan, and S. Li, “Service Function Chain Deployment Using Deep *Q* Learning and Tidal Mechanism,” *IEEE Internet Things J.*, vol. 11, no. 3, pp. 5401–5416, 2024.

[26] H. A. Shah and L. Zhao, “Multiagent Deep-Reinforcement-Learning-Based Virtual Resource Allocation Through Network Function Virtualization in Internet of Things,” *IEEE Internet Things J.*, vol. 8, no. 5, pp. 3410–3421, 2021.

[27] N. Toumi, M. Bagaa, and A. Ksentini, “Hierarchical Multi-Agent Deep Reinforcement Learning for SFC Placement on Multiple Domains,” in *2021 IEEE 46th Conference on Local Computer Networks (LCN)*, 2021, pp. 299–304.

[28] A. Pentelas, D. De Vleeschauwer, C.-Y. Chang, K. De Schepper, and P. Papadimitriou, “Deep Multi-Agent Reinforcement Learning With Minimal Cross-Agent Communication for SFC Partitioning,” *IEEE Access*, vol. 11, pp. 40384–40398, 2023.

[29] X. Wang, H. Xing, F. Song, S. Luo, P. Dai, and B. Zhao, “On Jointly Optimizing Partial Offloading and SFC Mapping: A Cooperative Dual-Agent Deep Reinforcement Learning Approach,” *IEEE Trans. Parallel Distrib. Syst.*, vol. 34, no. 8, pp. 2479–2497, 2023.

[30] K. Doan, M. Avgeris, A. Leivadeas, I. Lambadaris, and W. Shin, “Service Function Chaining in LEO Satellite Networks via Multi-Agent Reinforcement Learning,” in *GLOBECOM 2023 - 2023 IEEE Global Communications Conference*, 2023, pp. 7145–7150.

[31] D. Xiao, J. A. Zhang, X. Liu, Y. Qu, W. Ni, and R. P. Liu, “A Two-Stage GCN-Based Deep Reinforcement Learning Framework for SFC Embedding in Multi-Datacenter Networks,” *IEEE Trans. Netw. Serv. Manage.*, vol. 20, no. 4, pp. 4297–4312, 2023.

[32] S. Wang, H. Cao, L. Yang, S. Garg, G. Kaddoum, and M. Alrashoud, “GCN-Based Multi-Agent Deep Reinforcement Learning for Dynamic Service Function Chain Deployment in IoT,” *IEEE Trans. Consumer Electron.*, vol. 70, no. 3, pp. 6105–6118, 2024.

[33] P. Wei et al., “Reinforcement Learning-Empowered Mobile Edge Computing for 6G Edge Intelligence,” *IEEE Access*, vol. 10, pp. 65156–65192, 2022.

[34] J. Hao et al., “Exploration in Deep Reinforcement Learning: From Single-Agent to Multiagent Domain,” IEEE Transactions on Neural Networks and Learning Systems, vol. 35, no. 7, pp. 8762–8782, 2024.

[35] Z. Li, V. Chang, J. Ge, L. Pan, H. Hu, and B. Huang, “Energy-aware task offloading with deadline constraint in mobile edge computing,” *EURASIP Journal on Wireless Communications and Networking*, vol. 2021, no. 1, p. 56, 2021.

[36] H. Xia, W. Hao, K. Katsalis, Y. Kuno, D. Lee, J. Leung, M. Ju, F. Naim, J. Pieczerak, S. Schaller, T. Takahashi, I. Vancsa, and L. Willis, “Management and orchestration of the telco cloud: The role of nfv-mano and its added value,” European Telecommunications Standards Institute (ETSI), ETSI White Paper No. 67, May 2025, 1st ed.

[37] E. A. Hansen, D. S. Bernstein, and S. Zilberstein, “Dynamic programming for partially observable stochastic games,” in *Proceedings of the AAAI Conference on Artificial Intelligence*, vol. 4, 2004, pp. 709–715.

[38] W. Hao, M. Ju, A. Kumar, Y. Kuno, D. Lee, F. Naim, J. Triay, and H. Xia, “NFV evolution: Towards the Telco Cloud,” European Telecommunications Strandards Institute (ETSI), Tech. Rep. No. 65, March 2025, 1st ed.

[39] G. Papoudakis, F. Christianos, and S. Albrecht, "Agent Modelling under Partial Observability for Deep Reinforcement Learning," in *Advances in Neural Information Processing Systems*, vol. 34, pp. 19210–19222, 2021.
[40] A. Paszke, S. Gross, F. Massa, A. Lerer, J. Bradbury, G. Chanan, T. Killen, Z. Lin, N. Gimelshein, L. Antiga, A. Desmaison, A. Kopf, E. Yang, Z. DeVito, M. Raison, A. Tejani, S. Chilamkurthym, B. Steiner, L. Fang, J. Bai, and S. Chintala., "PyTorch: An Imperative Style, High-Performance Deep Learning Library," in *Advances in Neural Information Processing Systems*, vol. 32, 2019.
[41] M. Hessel J. Modayil, H. van Hasselt, T. Schaul, G. Ostrovski, W. Dabney, D. Horgan, B. Piot, M. G. Azar, and D. Silver, "Rainbow: Combining Improvements in Deep Reinforcement Learning," *Proceedings of the AAAI Conference on Artificial Intelligence*, vol. 32, no. 1, 2018, pp. 3215-3222.
[42] J. Dean and S. Ghemawat, "MapReduce: simplified data processing on large clusters," *Commun. ACM*, vol. 51, no. 1, pp. 107–113, Jan. 2008.
[43] S. Maheshwari, D. Raychaudhuri, I. Seskar, and F. Bronzino, "Scalability and Performance Evaluation of Edge Cloud Systems for Latency Constrained Applications," in *2018 IEEE/ACM Symposium on Edge Computing (SEC)*, 2018, pp. 286–299.
[44] B. Sonkoly, D. Haja, B. Németh. M. Szalay, J. Czentye, R. Szabó, R. Ullah, B.-S. Kim, and L. Toka, "Scalable edge cloud platforms for IoT services," *Journal of Network and Computer Applications*, vol. 170, p. 102785, 2020.

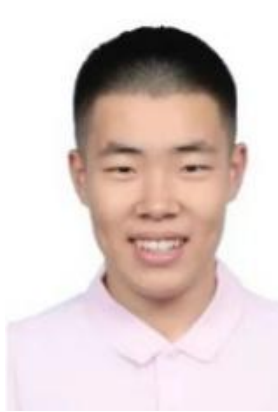

**Hanzhi Chang** received his B.S. degree from the Department of Cyber Science and Engineering, University of International Relations, Beijing, China, in 2023. He is currently pursuing for his M.S. degree in the Department of Cyber Science and Engineering, University of International Relations, Beijing, China. His research interests include network function virtualization, network resource orchestration and management, and reinforcement learning algorithms.

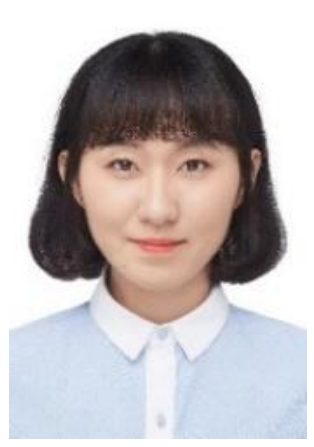

**Jing Bai** received the PhD degree in cyberspace security from Beijing Jiaotong University in 2023. She is currently a Lecturer at School of Cyber Science and Engineering, University of International Relations. Her interests include software trustworthiness analysis and cloud security.

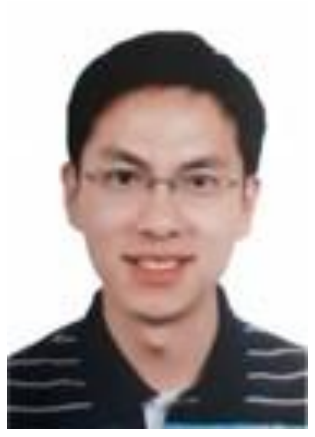

**Xin Tang** received the PhD degree in computer science from Beijing University of Posts and Telecommunications, Beijing, China, in 2015. He worked as a post-doctoral fellow in Department of Electronic Engineering at Tsinghua University, Beijing, China, from 2015 to 2017. He is currently an associate professor with the School of Cyber Science and Engineering, University of International Relations, Beijing, China. His current research interests are focused on secure cloud computing and reversible data hiding.

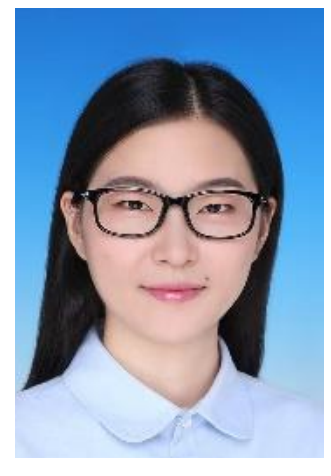

**Xiaomei Liu** received the Master's degree in computer technology from University of Chinese Academy of Sciences in 2018. She is currently an Engineer at School of Cyber Science and Engineering, University of International Relations. Her interests include cloud-edge computing and network security.

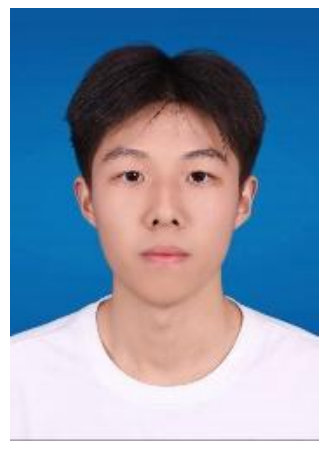

**Yiming Chen** is currently pursuing for his B.S. degree in the Department of Cyber Science and Engineering, University of International Relations, Beijing, China. His research interests include network function virtualization, network resource orchestration and management, and reinforcement learning algorithms.